\documentclass[12p,preprint]{aastex}

\shorttitle{O/H, Mg/H, Si/H, and Fe/H in \ion{H}{2} Regions}
\shortauthors{Peimbert and Peimbert}

\begin{document}

\title{On the O/H, Mg/H, Si/H and Fe/H Gas and Dust Abundance Ratios in Galactic and Extragalactic \ion{H}{2} Regions\footnotemark{}}

\author{Antonio Peimbert}
\affil{Instituto de Astronom\'\i a, Universidad Nacional Aut\'onoma de M\'exico, 
Apdo. Postal 70-264, M\'exico 04510 D.F., Mexico}
\email{antonio@astroscu.unam.mx}

\and
\author {Manuel Peimbert}
\affil {Instituto de Astronom\'\i a, Universidad Nacional Aut\'onoma de M\'exico,
Apdo. Postal 70-264, M\'exico 04510 D.F., Mexico} 
\email{peimbert@astroscu.unam.mx}

\begin{abstract}

We derive the Mg/H ratio in the Orion nebula and in 30 Doradus. We also derive the O/H and the 
Fe/O ratios in the extremely metal poor galaxy SBS~0335$-$052~E. We estimate the dust depletions of 
Mg, Si, and Fe in Galactic and extragalactic \ion{H}{2} regions. {From} these depletions we 
estimate the fraction of O atoms embedded in dust as a function of the O/H ratio. We find an 
increasing depletion of O with increasing O/H. The O depletion increases 
from about 0.08 dex, for the metal poorest \ion{H}{2} regions known, to about 0.12 dex, for 
metal rich \ion{H}{2} regions. This depletion has to be considered to compare nebular 
with stellar abundances.

\end {abstract}
 

\footnotetext{Based on observations collected at the European Southern
Observatory, Chile, proposals numbers: ESO 68.C-0149(A) and ESO 69.C-0203(A).}

\section{Introduction}

To determine the primordial helium abundance and to study the chemical evolution of 
galaxies it is necessary to derive the total $N$(O)/$N$(H) ratio in \ion{H}{2} regions; 
in this paper the abundances are given by number following the usual definitions 
X/H = 12 + log $N$(X)/$N$(H), X/O = log $N$(X)/$N$(O) = X/H $-$ O/H, and 
X$^{+i}$/H$^+$ = 12 + log $N$(X$^{+i}$)/$N$(H$^+$). To 
obtain the total O/H ratio in \ion{H}{2} regions it is necessary to add  the 
dust-phase to the gas-phase component of the O/H value. We presented elsewhere 
a preliminary discussion on the fraction of O trapped in dust grains, (Peimbert, A. 
\& Peimbert, M. 2010a). To find the fraction of O atoms embedded in dust grains we
will estimate the depletions of Mg, Si, and Fe, and we will assume that 
the O atoms trapped in dust belong to molecules that include Mg, Si, and Fe atoms.

There are three results in favor of dust presence in low metallicity \ion{H}{2} 
regions: a) Cannon et al. (2002) found significant enhancements of the 
$I$(H$\alpha$)/$I$(H$\beta$) value in some areas of I Zw 18 (one of the most metal-poor 
galaxies known) that are not due to 
underlying stellar absorption nor to the effect of collisional excitation of the 
Balmer lines, they conclude that the enhancements are more likely due to the presence 
of dust, b) Thuan et al. (1997) find that dust is clearly present in the extremely 
metal poor galaxy SBS~0335$-$052~E,  and c) the Si/O versus O/H diagram of \ion{H}{2} regions, 
where the Si/O ratio is almost independent of the O/H ratio and considerably smaller than the solar
ratio implying that a substantial fraction of Si is embedded in dust grains
(Garnett et al. 1995).

There are two arguments in favor of significantly lower depletion 
fractions of heavy elements into dust in metal poor 
\ion{H}{2} regions: a) Izotov et al. (2006) find a slight increase of Ne/O with 
increasing metallicity, which they interpret as due to a moderate depletion of O 
onto grains in the most metal-rich galaxies, they conclude that this depletion 
corresponds to $\sim$ 20\% of oxygen locked in the dust grains in the 
highest-metallicity \ion{H}{2}  regions of their sample, while no significant 
depletion would be present in the \ion{H}{2} regions with lower metallicity, 
and b) Rodr\'{\i}guez \& Rubin (2005) have estimated the fraction of Fe atoms 
in Galactic and extragalactic \ion{H}{2} regions in the gaseous phase, and find 
that this fraction decreases with metallicity. The Fe
depletions derived for the different objects define a trend of higher depletion 
at higher metallicities; while the Galactic \ion{H}{2} regions show less than  
5\% of their Fe atoms in the gas phase, the low metallicity extragalactic regions (LMC 30 Doradus, 
SMC N88A, and SBS~0335$-$052~E) have somewhat lower fractions of Fe  embedded in dust 
grains. Izotov et al. (2006) have produced a Fe/O versus O/H diagram that confirms
the results by Rodr\'{\i}guez \& Rubin.

To estimate the dust fraction as a function of O/H we need to have good gaseous 
abundance determinations for a relatively O-rich  \ion{H}{2} region; for this  
purpose we will use the the Orion 
nebula, its gaseous abundances derived by Esteban et al. (2004),
and the result by Mesa-Delgado et al. (2009) that from three different methods 
have estimated that the fraction of O atoms trapped in dust in the Orion nebula 
amounts to 0.12 $\pm$ 0.03 dex. We will compare the Orion nebula abundances with 
the protosolar abundances by Asplund et al. (2009), and the solar vicinity B stars 
abundances by Przybilla et al. (2008).

Very O poor \ion{H}{2} regions are also needed to find the fraction of O embedded in dust. 
Therefore we decided to compute the O/H and Fe/H abundances of SBS~0335$-$052~E, and 
to consider results for other \ion{H}{2} regions available in the literature.

In sections 2 and 3 we determine the Mg/H value, adopt a set of O/H, Si/H, and Fe/H 
gaseous values and estimate the fractions of O, Mg, Si and 
Fe embedded in dust grains for the Orion nebula and for 30 Doradus, respectively. 
In section 4 we compute the O/H and Fe/H abundances 
for  SBS~0335$-$052~E based on: the O/H abundances derived from recombination lines, 
considering the density dependence of the \ion{O}{2} recombination lines, and 
the density dependence of the [\ion{Fe}{3}] lines computed by Keenan et al. (2001). 
In section 5 we make use 
of the Si/O ratios for 5 extragalactic \ion{H}{2} regions obtained by Garnett 
et al. (1995). In section 6 we compile  
Fe/O ratios from the literature for Galactic and extragalactic \ion{H}{2} regions 
and estimate the Fe dust fraction as a function of O/H. In section 7 we  discuss 
the observational results and estimate 
the fraction of O atoms trapped in dust grains as a function of O/H, and the 
conclusions are presented in section 8.

\section{The Orion nebula}

In Table 1 we present the solar neighborhood and Orion Nebula abundances that will be 
used as reference values to estimate the fraction of a given element in the dust phase. 
In column 2 we present the protosolar abundances (Asplund et al. 2009). In column 3 we 
present the solar neighborhood abundances by correcting the protosolar abundances due to 
the chemical evolution of the Galaxy since the Sun was formed; based on the Galactic 
chemical evolution models of Carigi (1996), Chiappini et al. (2003), and Carigi 
\& Peimbert (2008, 2010) we added 0.09, 0.09, 0.13, and 0.20 dex to the protosolar values of O, 
Mg, Si, and Fe respectively. In column 4 we present the chemical abundances 
derived by Przybilla et al. (2008) from 6 B stars of the solar vicinity; Sim\'on-D\'{\i}az (2010)
has recently derived the O and Si abundances for 13 B stars from the Orion region, his results 
are in very good agreement with those by Przybilla et al.; note that B stars abundances do not 
need correction for Galactic chemical evolution since these stars are only a few million years old.  
In column 5 we present our adopted abundances for the ISM of the solar vicinity; 
we obtained these values from a simple average of columns 3 and 4. In column 6 we present the 
expected total abundances for the Orion Nebula; we obtained these values by considering the 
different galactocentric distances of the solar vicinity and the Orion Nebula and subtracting 
to the O, Mg, Si, and Fe solar neighborhood abundances 0.02, 0.02, 0.03, and 0.04 dex 
respectively to account for the presence of Galactic abundance gradients; the Galactic gradients 
are obtained from the Galactic chemical evolution models by Carigi, Chiappini and their collaborators; 
it is worth mentioning that the expected total, gas plus dust, O/H value for the Orion nebula is equal to the total derived O/H value from observations
by Mesa-Delgado et al. (2009) (see below). 
In column 7 we present the gaseous abundances of the Orion nebula; the O/H 
abundance comes from Esteban et al. (2004), the Fe/H abundance from Esteban et al. (1998), 
and the Mg/H and Si/H abundances from this paper (see below).

\begin{deluxetable}{lr@{$\pm$}lr@{$\pm$}lr@{$\pm$}lr@{$\pm$}lr@{$\pm$}lr@{$\pm$}lr@{}lr@{}l}
\rotate
\tablecaption{Gas and Dust Abundances\tablenotemark{a}~\null~\null~for the Solar Vicinity and the Orion Nebula
\label{tab:Orion}}
\tablehead{
\colhead{} &
\multicolumn{2}{c}{} &
\multicolumn{2}{c}{Protosolar} &
\multicolumn{2}{c}{B stars} &
\multicolumn{2}{c}{ISM} &
\multicolumn{2}{c}{Orion} &
\multicolumn{2}{c}{Orion} &
\multicolumn{2}{c}{Orion} & 
\multicolumn{2}{c}{Orion} \\
\colhead{Element} &
\multicolumn{2}{c}{Protosolar\tablenotemark{b}} &
\multicolumn{2}{c}{+ GCE\tablenotemark{bc}} &
\multicolumn{2}{c}{solar vicinity\tablenotemark{d}} &
\multicolumn{2}{c}{solar vicinity\tablenotemark{e}} &
\multicolumn{2}{c}{g + d\tablenotemark{f}} &
\multicolumn{2}{c}{gas\tablenotemark{g}} &
\multicolumn{2}{c}{gas fraction} & 
\multicolumn{2}{c}{dust fraction} 
}
\startdata

O  & 8.73&0.05 & 8.82&0.06 & 8.76&0.03 & 8.79&0.03 &  8.77&0.03  & 8.65&0.03 & (76&$\pm$5)\% & (24&$\pm$5)\% \\
Mg & 7.64&0.04 & 7.73&0.07 & 7.56&0.05 & 7.64&0.08 & (7.62&0.05) & 6.56&0.16 &  (9&$\pm$3)\% & (91&$\pm$3)\% \\
Si & 7.55&0.04 & 7.68&0.09 & 7.50&0.02 & 7.59&0.09 & (7.56&0.04) & 6.94&0.12 & (24&$\pm$7)\% & (76&$\pm$7)\% \\
Fe & 7.54&0.04 & 7.74&0.14 & 7.44&0.04 & 7.59&0.15 & (7.55&0.04) & 6.07&0.20 &  (3&${+2\atop-1}$)\% 
                                                                                      & (97&${+1\atop-2}$)\% \\

\enddata
\tablenotetext{a}{Abundances in units of 12 + log $N$(X)/N$(H)$.}
\tablenotetext{b}{Protosolar values from Asplund et al. (2009).}
\tablenotetext{c}{Galactic chemical evolution correction from Carigi (1996), Chiappini et al. (2003), and
Carigi \& Peimbert (2008, 2010), see text.}
\tablenotetext{d}{Solar vicinity B stars from Przybilla et al. (2008).}
\tablenotetext{e}{Adopted solar vicinity ISM abundances, average from B stars and protosolar values
corrected for Galactic chemical evolution.}
\tablenotetext{f}{O from Esteban et al. (2004) including the dust correction from 
Mesa-Delgado et al. (2009); values in parenthesis come from the ISM solar vicinity values presented in the fifth
column and a small correction due to the larger galactocentric distance of Orion relative to the Sun
predicted by the Galactic chemical evolution models mentioned above.}
\tablenotetext{g}{O from Esteban et al. (2004), Fe from Esteban et al. (1998), Mg and Si from this paper.}
\end{deluxetable}

Baldwin et al. (1991) estimated an upper limit to the gaseous Mg/H value of 6.58 dex from an upper limit to the \ion{Mg}{2} $\lambda$2800/H$\beta$ line intensity ratio obtained by 
Perinotto \& Patriarchi (1980). {From} the same raw data of the observations for Orion by Esteban et al. (2004), we have estimated the gaseous Mg abundance in this region. From these data, we have measured the following line ratios: $I$(4481.22, \ion{Mg}{2})/$I$(H$\beta$) = 
(3.6$\pm$1.6) x 10$^{-5}$ and 
$I$(4482.17, \ion{Fe}{1}])/$I$(H $\beta$) = (8.6$\pm$2.5) x 10$^{-5}$. As we can see in Figure 1, high resolution spectra are needed to separate the \ion{Mg}{2} line from the 4482.2 line, that is 
probably due to the \ion{Fe}{1}] (a$^5$-z$^7$F$^o$).  Under the assumption that the \ion{Mg}{2} 3d-4f 
$\lambda$4481 line has an effective recombination coefficient equal to that of the  
\ion{C}{2} 3d-4f $\lambda$4267 line (Liu et al. 2004), from the \ion{Mg}{2}/H$\beta$ intensity ratio and the \ion{C}{2} line effective recombination coefficient we derive a gaseous 
Mg$^{++}$/H$^+$ value of 6.54 $\pm$ 0.16 dex. By adding 0.02 dex due to 
the presence of Mg${^+}$ (Baldwin et al. 1991), we obtain a total Mg/H gaseous component 
of 6.56 $\pm$ 0.16 dex. Our Mg/H ratio is in agreement with the upper limit  obtained by Baldwin et al., this ratio together with the reference stellar value of 
Table 1, implies that (9$\pm$3)\% of the Mg atoms are in  gaseous form and (91$\pm$3)\%  are embedded in dust.

\begin{figure}
\plotone{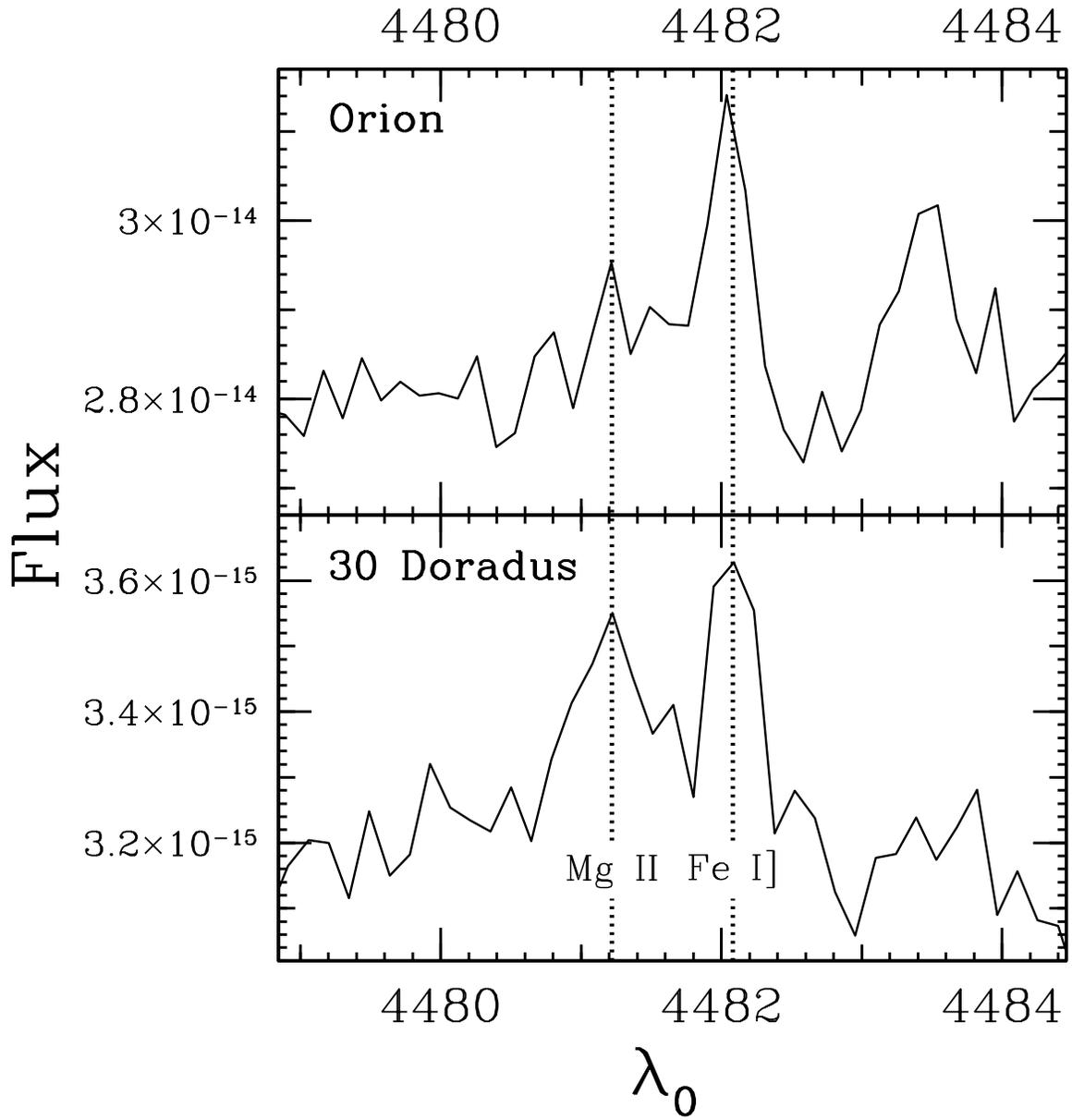}
\caption{Spectra of the Orion nebula and the 30 Doradus nebula that show the 
$\lambda\lambda$ 4481.22, \ion{Mg}{2} and 4482.17, \ion{Fe}{1}] emission lines, the rest wavelength is given
in Angstroms and the flux in ergs cm$^{-2}$ s$^{-1}$ A$^{-1}$.
\label{fig:MgII lines}}
\end{figure}

Garnett et al. (1995), obtained UV observations of the Orion nebula with the HST, and found that 
Si/C $= -1.46 \pm 0.10 $ dex. From the C/O value of $-0.25$ dex derived by
Esteban et al. (2004) and the previous result we find that the gaseous Si/O value amounts to $-1.71$ dex. By considering that 0.12 dex of O is tied up in dust grains it follows that the gaseous 
Si to the total (gas + dust) O abundance is equal 
to $-1.83$ dex, and consequently that Si/H $= 6.94 \pm 0.12$ dex. This 
number together with the reference Si/H value implies that the fraction of Si atoms 
tied up in dust grains amounts to 76\% in the Orion nebula (see Table 1). Similarly 
Esteban et al. (1998) obtained that the gaseous Fe/H $= 6.07 \pm 0.20$ dex, this result together
with the Fe/H reference value implies that about 97 \% of the Fe atoms are tied up in dust grains.

The total O value for the Orion nebula presented in Table 1, O/H = 8.77 dex, was 
obtained from 2 different methods: a) By adding to the gaseous value derived by 
Esteban et al. (2004) for $t^2$ = 0.022 the dust correction of 0.12 dex estimated 
by Mesa-Delgado et al. (2009), where $t^2$ is
the mean square temperature variation (Peimbert 1967). And b) by using the same 
procedure for O than the one used for the other elements of Table 1.

The use of the $t^2$ formalism  presented by Peimbert (1967)
and Peimbert \& Costero (1969) will be used throughout this paper. 
As mentioned before the use of $t^2$ = 0.022 gives O/H = 8.65, a reasonable value for the estimated
oxygen depletion. On the other hand under the simpler assumption, that \ion{H}{2} regions
have uniform temperature ($t^2$ = 0.000), Deharveng et al. (2000), Pilyugin, 
Ferrini, \& Shkvarun (2003), and Esteban et al. (2004) obtain for the Orion nebula O/H 
values of 8.51, 8.49, and 8.51 dex respectively, a difference of 0.27 dex with respect 
to the total oxygen value presented in Table 1. A difference too large to be explained by the 
depletion of oxygen into dust grains.
It should be noted that: of the well observed \ion{H}{2} regions  
the Orion nebula presents one of the lowest 
$t^2$ values (e.g. Esteban et al. 2002, 2005, 2009, Peimbert 2003, Peimbert et al. 2007, 
Garc\'{\i}a-Rojas et al. 2007), therefore in general we expect larger abundance 
differences than in Orion between the $T$(4363/5007) method and the method with
$t^2$ values different from zero.

We will assume that the fraction of O atoms tied up in dust grains of \ion{H}{2} 
regions is proportional to the fraction of atoms of Mg, Si, and Fe tied up in dust 
grains of each nebula and that the constant of proportionality will be given by 
the Orion nebula results. This assumption is a first approximation to this 
problem and is based on the following ideas: a) as far as molecules --- that survive 
as part of dust grains in an \ion{H}{2} region --- go, the elements 
that most frequently bind with O atoms are Mg, Si, and Fe; b) Na, Al, Ca, Cr, Mn, and Ni are less abundant 
than the previous three elements, and most of them have similar behaviors to those of 
either Mg, Si, or Fe; c) the O atoms tied up by Mg, Si, and Fe are present in 
molecules like: [(Mg,Fe)$_2$SiO$_4$] and [(Mg,Fe)SiO$_3$] (Ossenkopf, Henning, \& 
Mathis 1992); Fe$_2$O$_3$ and Fe$_3$O$_4$ (Nuth \& Hecht 1990); and MgO (Fadeyev 1988); 
and d) in these molecules about one O atom is attached to a Mg, Si, or Fe atom.

For the Orion Nebula we know that $N({\rm O}_{dust})/N({\rm O}_{total})=0.241\pm0.060$.
{From} this result, the previous assumption, and the values of Table 1 we obtain that:
\begin{equation}
\frac{N({\rm O})_{dust}}{N({\rm O})_{total}} =
(0.241\pm0.060) \frac{\left[ \frac{N({\rm Mg})_{dust}}{N({\rm Mg})_{total}} + 
                             \frac{N({\rm Si})_{dust}}{N({\rm Si})_{total}} + 
                             \frac{N({\rm Fe})_{dust}}{N({\rm Fe})_{total}} \right] }{2.64}
.
\label{edust}
\end{equation}
The error bars from equation 1 correspond to a systematic uncertainty that depends on the exact 
O depletion of the Orion Nebula; we will ignore this systematic error for most of the paper, 
and will look into it at the end of section 7.

We will estimate the ${\rm O}_{total}$ for other \ion{H}{2} regions from equation (1) and the 
next equation:
\begin{equation}
\left[\frac{N({\rm O})}{N({\rm H})}\right]_{total} =
\frac{1}{1-0.2739} 
\left(\left[\frac{N({\rm O})}{N({\rm H})}\right]_{gas} -  0.0913 \left\{
                      \frac{\left[\frac{N({\rm Mg})}{N({\rm H})}\right]_{gas}}
                            {\left[\frac{N({\rm Mg})}{N({\rm O})}\right]_{ref}} + 
                      \frac{\left[\frac{N({\rm Si})}{N({\rm H})}\right]_{gas}}
                            {\left[\frac{N({\rm Si})}{N({\rm O})}\right]_{ref}} + 
                      \frac{\left[\frac{N({\rm Fe})}{N({\rm H})}\right]_{gas}}
                            {\left[\frac{N({\rm Fe})}{N({\rm O})}\right]_{ref}}
                                                                         \right\}\right).
\label{egas}
\end{equation}

Jenkins (2009) finds for some lines of sight of the ISM, that include mainly neutral material,
that the O depletion  reaches values in the 0.2 to 0.3 dex range. He concludes
that O molecules that include Mg, Si, and Fe are not enough to account for these depletions, and suggests 
that ices might also be present. Equations (1) and (2) were calibrated with the Orion nebula, where
according to Mesa-Delgado et al. (2009) the O depletion derived from the assumption that O is trapped
by molecules, that include only Mg, Si, and Fe, agrees with: a) the O depletion estimated from the comparison
of the stellar abundances of the Orion region with the nebular abundances; and b) with the O pre-shock 
nebular abundances and post shock O abundances associated with the H-H object 202, in this object 
after the shockwave passes dust grains are destroyed and most of the 
Fe trapped in grains is returned to the gaseous phase, they assume that also most of the O is returned 
to the ISM. Since the Orion 
nebula is one of the \ion{H}{2} regions with the highest Fe depletion we consider unlikely that 
the O depletions of \ion{H}{2} regions with lower Fe depletions are higher than in the Orion 
nebula, therefore we conclude that ices and O$_2$ molecules are practically absent in \ion{H}{2} regions.

\section{The 30 Doradus nebula}

To obtain the fraction of O, Mg, Si and Fe in the dust phase of 30 Doradus we will 
follow the same procedure as for the Orion nebula.

Based on the raw data presented by Peimbert (2003) we derived the these two line 
ratios: $I$(4481.22, \ion{Mg}{2})/$I$(H$\beta$) = (1.11$\pm$0.28) x 10$^{-4}$ and 
$I$(4482.17, \ion{Fe}{1}])/$I$(H$\beta$) = (1.06$\pm$0.27) x 10$^{-4}$, the observed  \ion{Mg}{2} and \ion{Fe}{1}]
lines are presented in Figure 1. From the 4481 \ion{Mg}{2} line intensity 
we obtained that Mg$^{++}$/H$^+$ = 7.06$\pm$0.12 dex. The Mg/H total gaseous value is presented 
in Table 2, due to the considerably higher degree of ionization of 30 Doradus relative 
to the Orion nebula we neglected the contribution to this value due to Mg$^{+}$ .

Tsamis and P\'equignot (2005), from the observations of $\lambda\lambda$ 4561 and 
4571 of \ion{Mg}{1}] by Peimbert (2003), also derived the Mg/H value for 30 Doradus and it 
is presented in Table 2. To obtain a representative average of the two sets of values, we
normalized them by matching the two O/H values and consequently
we added 0.05 dex to the O/H,  Mg/H, Si/H, and Fe/H gaseous determinations by Tsamis 
and P\'equignot and averaged them with the 
O and Fe determinations by Peimbert (2003), the Si determination by Garnett et al. (1995), and the Mg determination presented in this paper (see Table 2).

\begin{deluxetable}{lr@{$\pm$}lr@{$\pm$}lr@{$\pm$}lr@{$\pm$}lr@{}lr@{}l}
\tablecaption{Gas and Dust Abundances\tablenotemark{a}~\null~\null~for 30 Doradus
\label{tab:30 Doradus}}
\tablewidth{0pt}
\tablehead{
\colhead{Element}       &
\multicolumn{2}{c}{gas\tablenotemark{b}} &
\multicolumn{2}{c}{gas\tablenotemark{c}}       &
\multicolumn{2}{c}{gas\tablenotemark{d}}       &
\multicolumn{2}{c}{g + d\tablenotemark{e}}       &
\multicolumn{2}{c}{gas fraction}    &    
\multicolumn{2}{c}{dust fraction}      
}
\startdata

O    & 8.50&0.05 & 8.45&0.05 & 8.50&0.05 &  8.61&0.03   & (78&$\pm$5)\%         & (22&$\pm$5)\%         \\
Mg   & 7.06&0.10 & 6.69&0.16 & 6.90&0.13 & (7.46&0.05)  & (28&$\pm$9)\%         & (72&$\pm$9)\%         \\
Si   & 6.80&0.27 & 6.50&0.27 & 6.68&0.27 & (7.40&0.04)  & (19&${+16\atop-9}$)\% & (81&${+16\atop-9}$)\% \\
Fe   & 6.39&0.20 & 6.12&0.20 & 6.28&0.20 & (7.39&0.04)  &  (8&${+4\atop-3}$)\%  & (92&${+4\atop-3}$)\%  \\

\enddata
\tablenotetext{a}{Abundances in units of 12 + log $N$(X)/N$(H)$.}
\tablenotetext{b}{Gaseous abundances for $t^2$ = 0.033. O and Fe from Peimbert (2003). Si
from Garnett et al. (1995). Mg from this paper.}
\tablenotetext{c}{Gaseous abundances for the inhomogeneous model by Tsamis \& P\'equignot (2005).}
\tablenotetext{d}{Adopted gaseous abundances from the average of the previous two 
columns, where all the values of the previous column were increased by 0.05 dex (see text).} 
\tablenotetext{e}{Reference gas plus dust abundances, given in parenthesis, come from the 
X/O values of Table 1. The total O abundance was derived from equation (2).}
\end{deluxetable}

We adopted for 30 Doradus the Mg/O, Si/O, and Fe/O reference values of the Orion nebula.
{From} these values, the observed gas phase Mg/H, Si/H, and Fe/H 
abundances, and equation 2, we derived the O/H total value.
By comparing the total gaseous abundances with the reference values 
presented in Table 2 we derived the fractions of the Mg, Si, and Fe atoms in the gas 
phase and the dust phase. The 30 Doradus Mg and Fe fractions are slightly lower than those of 
Orion, while Si is nearly constant; this result gives us an initial hint that the fraction 
of heavy elements trapped in dust diminishes at lower metallicities.

\section{The Fe/H and O/H ratios in SBS~0335$-$052~E}

It is important to obtain the Fe/O and the O/H ratio of the metal poorest \ion{H}{2} regions 
to try to find out if a substantial fraction of Fe is tied up in dust grains and consequently 
also a non-negligible amount of O atoms. To reach this objective we will compute the abundances 
for SBS~0335$-$052~E.

\subsection{The Fe/H ratio}

Recently Izotov et al. (2009) derived the Fe/H value for SBS~0335$-$052~E from
observations obtained with UVES and FORS.  On the one hand  their results based on the 
$\lambda$ 4987 [\ion{Fe}{3}] line need to be corrected because with the FORS resolution 
the line is a blend of $\lambda\lambda$ 4986+4987, and the contribution of $\lambda$ 4986 
has to be taken into account due to the large intensity of this line at low densities. We 
have obtained a density value of $108 \pm 12$~cm$^{-3}$ for SBS~0335$-$052E from Figure 2 
--- where we present the I(4986)/I(4658) and I(4987)/I(4658) ratios as a function of density 
--- and the [\ion{Fe}{3}] line intensities of this object measured by Izotov et al (2009). 
On the other hand, the Fe values derived from 4658 are almost density independent. From the 
$I$(4658) observations in SBS~0335$-$052~E Izotov et al. (2009) obtain an 
Fe/O = $-1.36 \pm 0.03$ dex in the gas phase. 

To calculate the total fraction of Fe in dust grains, we first need to obtain the Fe$_{gas}$/O$_{total}$ value
and then compare it to the reference value. To correct for the fraction of O in dust grains we will 
use equation 2 and the reference values derived for the Orion nebula. The gaseous Si/O ratio comes 
from Garnett et al. (1995) that obtain Si/O = $-1.72 \pm 0.20$ dex. 
We will assume that the value of the Mg term of equation 2 lies between the values of the 
Si and Fe terms, we will discuss this point further in section 7. From the previous considerations 
we derive a correction of $0.08\pm0.01$ dex for the O abundance.
Therefore we obtain an Fe$_{gas}$/O$_{total}$ of $-1.44 \pm 0.03$ dex. Taking as reference value the solar
vicinity ISM abundances, the Fe/O ratio amounts to $-1.20 \pm 0.15$ dex, 
and we obtain that the fraction of Fe in gaseous form amounts to $(58\pm20)$\% and that the rest is trapped 
in dust grains.

\begin{figure}
\plotone{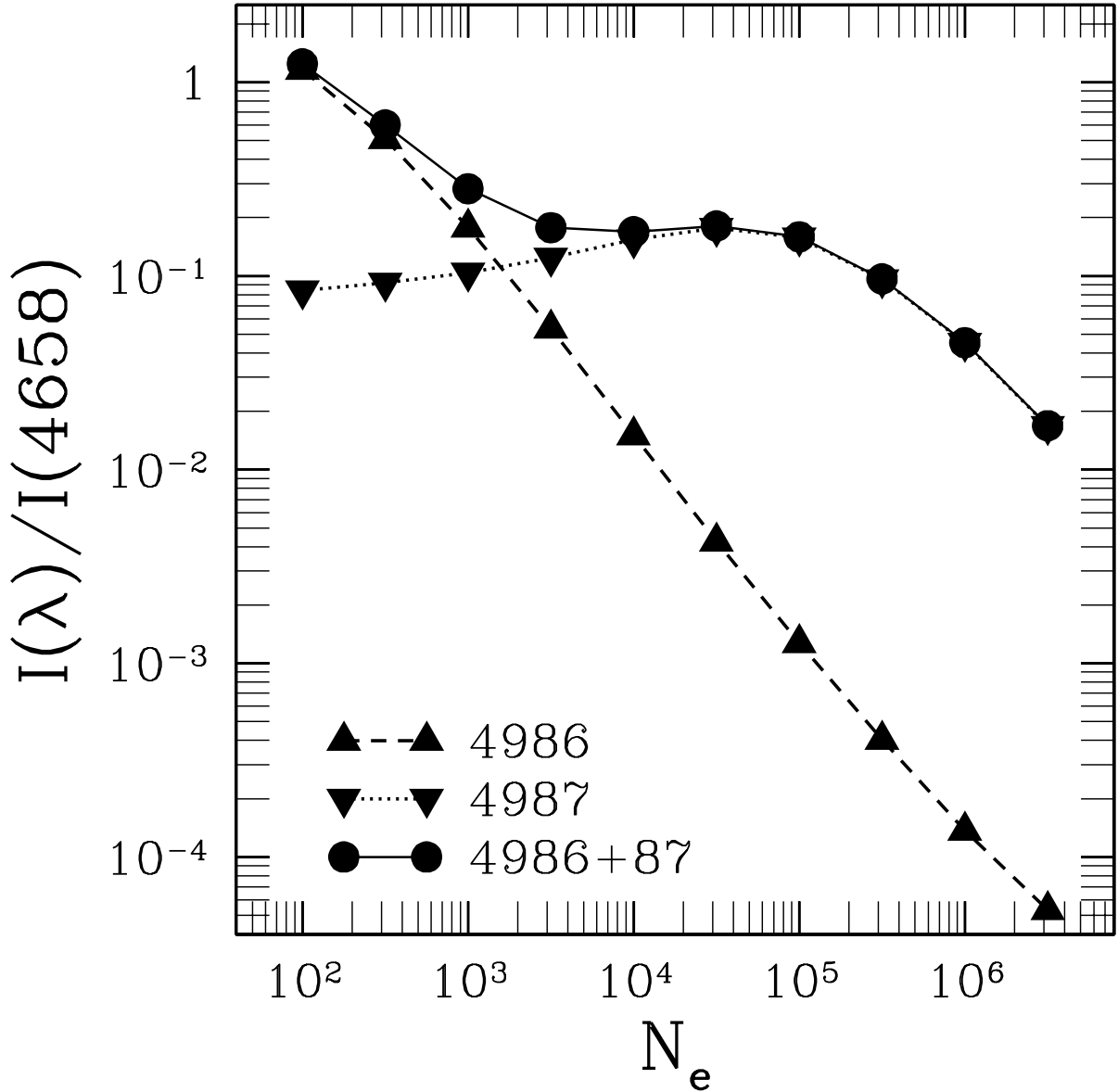}
\caption{[\ion{Fe}{3}] line intensity ratios. $I(\lambda)/I(4658)$ versus $N_e$ for $T_e$ = 15, 000 K, from the computations by Keenan et al. (2001).
\label{fig:FeIII}}
\end{figure}

\subsection{The O/H ratio}

Due to the presence of temperature inhomogeneities the abundances derived from 
$T$(4363/5007) and the [\ion{O}{3}] lines are only  lower limits to the O/H value. 
Therefore it is very important to obtain the O abundances from recombination 
lines that are independent of the temperature structure. Moreover the combination 
of the forbidden and permitted lines of O allow us to estimate the value of $t^2$.

To derive an accurate O/H value it is also necessary to consider the density
dependence of the \ion{O}{2} lines. 
The sum of the intensities of all the lines of 
the \ion{O}{2} multiplet 1, $I$(total), has a temperature and density dependence 
that is proportional to $T^{-0.9}N^{2}$, similar to the temperature and density dependence 
of the \ion{H}{1} lines. However the fraction of the intensity due to each one of the 
lines of the \ion{O}{2} multiplet 1 varies when $N_e<10000$ cm$^{-3}$. In Figure 3
we present the density dependence of the eight \ion{O}{2} lines of multiplet 1 that were obtained 
empirically from observations of planetary nebulae and \ion{H}{2} regions 
(Ruiz et al. 2003, Peimbert \& Peimbert 2005a, and Peimbert et al. 2005b). Bastin 
and Storey (2006) from atomic physics computations obtain a similar behavior for 
the $I$(4649)/$I$(total) versus density relation to that derived empirically from observations.

\begin{figure}
\plotone{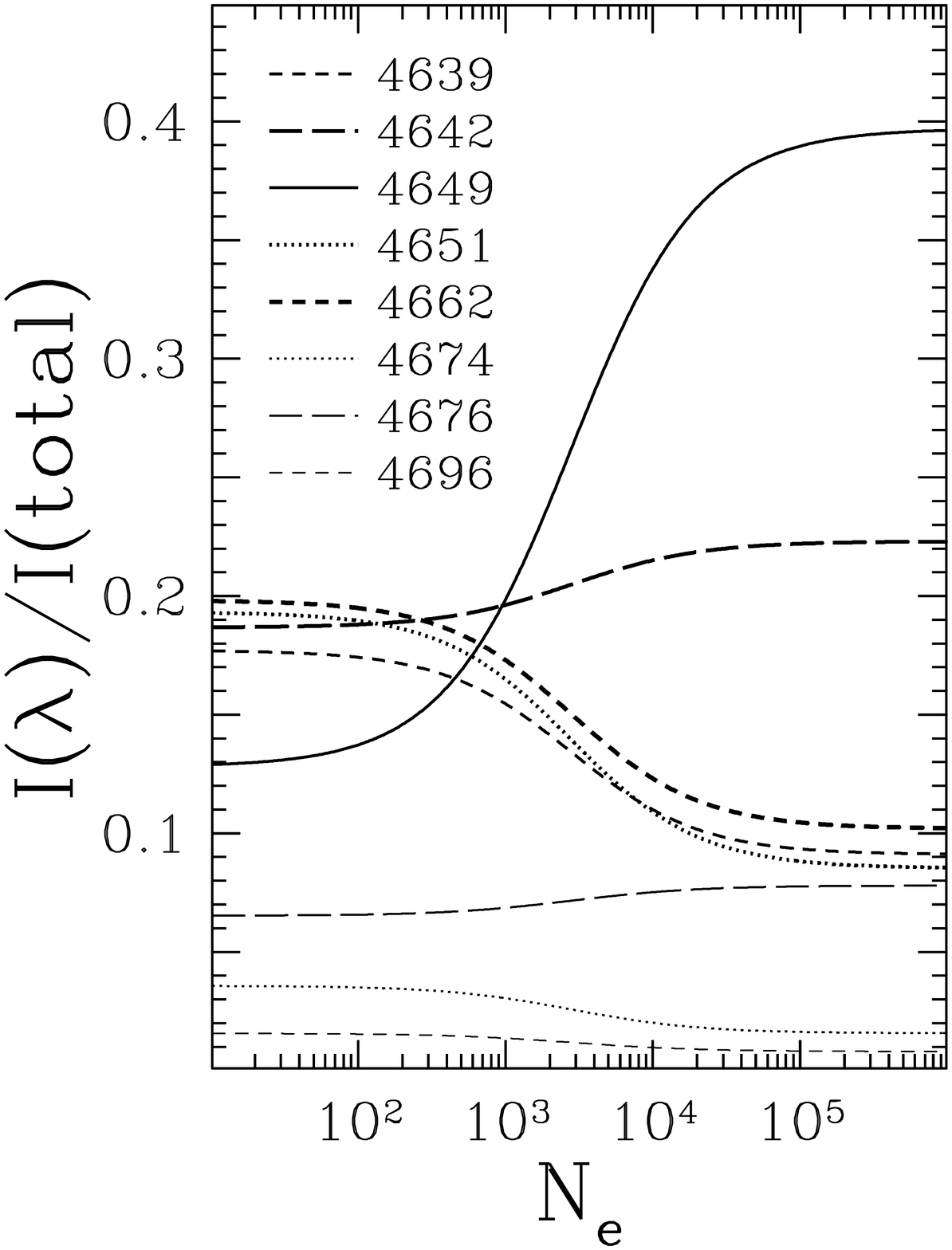}
\caption{$I(\lambda$)/$I$(total) versus $N_e$ for the \ion{O}{2} line intensities,
based on the equations derived from observations by:  Ruiz et al. (2003), 
Peimbert \& Peimbert (2005a), and Peimbert et al. (2005b).
\label{fig:OII}}
\end{figure}

To obtain abundances from the O recombination lines it is necessary to take into 
account the possible blend of the \ion{O}{2} lines when the resolution 
is not high enough. {From} the equations by Ruiz et al. (2003), 
Peimbert \& Peimbert (2005a), and Peimbert et al. (2005b), for those cases where 
the resolution is of the order of two to three angstroms, we recommend these equations:
\begin{equation}
\left[ \frac{I(4639+42)}{I({\rm total})} \right]_{obs} =
0.315 + \frac{0.051\pm0.008}{ \left[ 1 + N_e({\rm FL})/1325 \right] },
\label{e4639}
\end{equation}
\begin{equation}
\left[ \frac{I(4649+51)}{I({\rm total})} \right]_{obs} =
0.482 - \frac{0.170\pm0.011}{ \left[ 1 + N_e({\rm FL})/1325 \right] },
\label{e4649}
\end{equation}
and
\begin{equation}
\left[ \frac{I(4674+76)}{I({\rm total})} \right]_{obs} =
0.093 + \frac{0.007\pm0.003}{ \left[ 1 + N_e({\rm FL})/1325 \right] }.
\label{e4674}
\end{equation}
In Figure 4 we present what fraction of the total intensity is present in each of 
these blends as a function of density.

\begin{figure}
\plotone{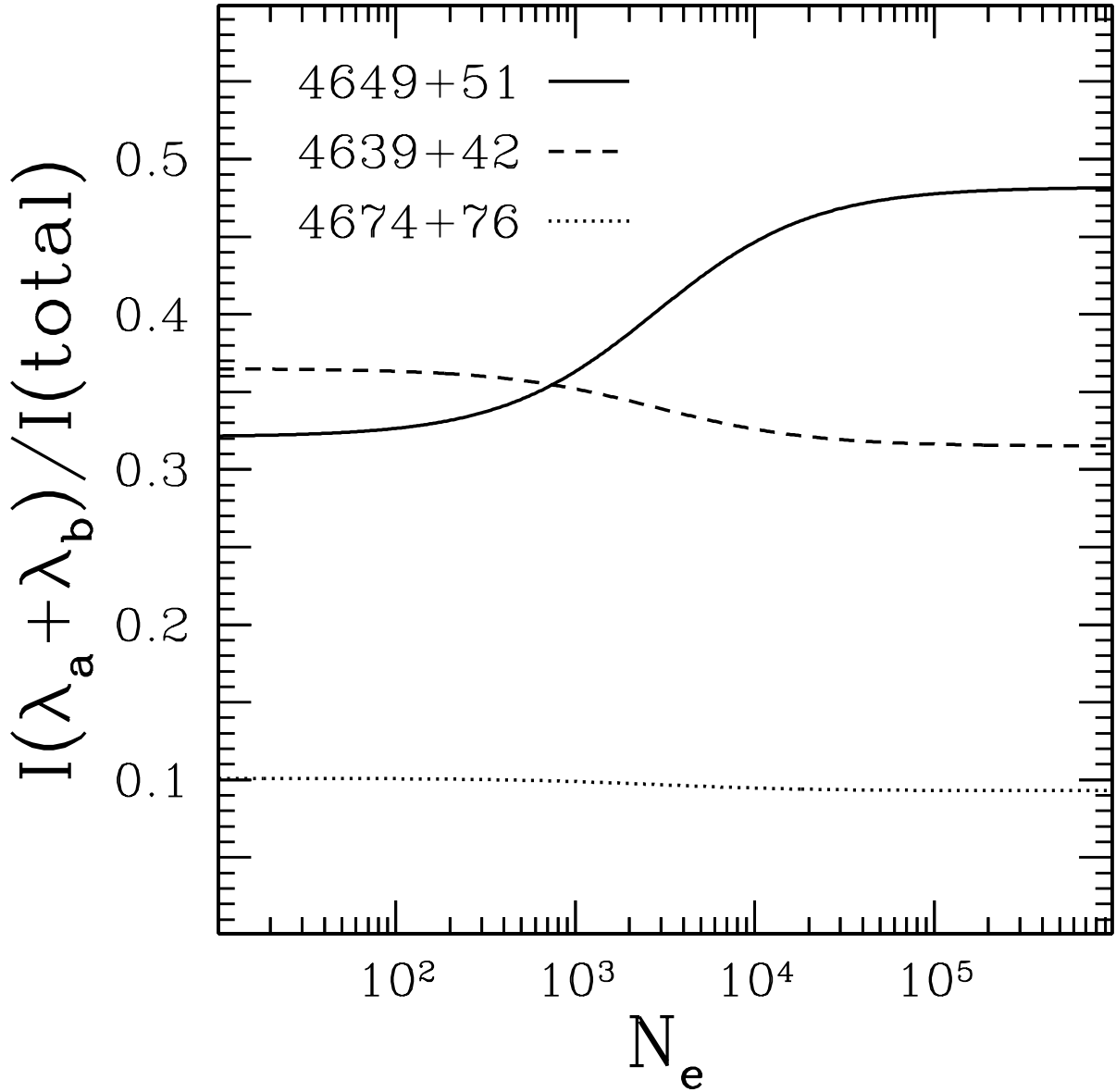}
\caption{$I(\lambda_a + \lambda_b)$/$I$(total) versus $N_e$, based on equations 3-5.
\label{fig:CarbonOHLWY}}
\end{figure}

Due to the importance and the faintness of the \ion{O}{2} recombination lines we decided to
reduce the FORS raw data for SBS~0335$-$052~E present in the VLT archive, 
an independent reduction of this data was made by Izotov et al. (2009). We obtain that  
$I$ (4649 + 51)/$I$(H$\beta)$ amounts to $0.0132 \pm 0.0074 $, in addition we obtain that
$I$(4639 + 42)/$I$(H$\beta$) amounts to 
$0.0085 \pm 0.0074$. From our two measurements we find that 
$I$(4639 + 4641 + 4649 + 51)/$I$(H$\beta) = 0.0217 \pm 0.0104$. We have subtracted
to $I$(H$\beta)$ the contribution due to collisional excitation, that according 
to Peimbert et al. (2007) amounts to 6.6\%, and obtain that 
$I$(4639 + 4641 + 4649 + 51)/$I$(H$\beta)_{rec} = 0.0231 \pm 0.0111$.
Based on the $I$(4639 + 42)/$I$(4649 + 51) line ratio and the very low N/O ratio in this object we 
consider that the feature at $\lambda\lambda 4639 + 4642$ is due to the \ion{O}{2} lines and not to 
the 4642 \ion{N}{3} line, contrary to the suggestion by Izotov et al. (2009).

>From our measurements, equations (3) and (4), the recombination coefficients 
of \ion{O}{2} by Storey (1994), and the \ion{H}{1} recombination coefficients by Storey and 
Hummer (1995) we obtain an abundance of O$^{++}$/H$^+$ = $7.36{+0.17 \atop -0.28}$ dex. 
The total O/H value amounts to $7.41{+0.17 \atop -0.28}$
dex and includes the O$^+$ contribution that amounts to 0.05 dex 
(Luridiana et al. 2003, Izotov et al. 2009). Our O/H value can be compared 
with the values for $t^2 = 0.00$  derived
by Peimbert et al. (2007) and Izotov et al. (2009) that amount to $7.31 \pm 0.04$ dex and 
$7.23 \pm 0.01$ dex respectively, and also with the value derived by Peimbert 
et al. for $t^2 = 0.04$ that amounts to $7.47 \pm 0.04$ dex. In Table 3 we present 
the value for $t^2 = 0.00$ by Peimbert et al. (2007), 
we also present an the O/H value derived by Peimbert et al. for $t^2 = 0.04$.

>From the FORS raw data for SBS~0335$-$052~E Izotov et al. (2009)  obtain that 
$I$(4649 + 51)/$I$(H$\beta$) = $0.02 \pm 0.01$. From this line intensity ratio they obtain that
O$^{++}$/H$^+$ = $7.18 {+0.17 \atop -0.30}$ dex, and by adding the O$^{+}$ contribution 
a total O/H = $7.23 {+0.17 \atop -0.30}$ dex. From their published line intensity, the recombination coefficients by 
Storey (1994), and Storey and Hummer (1995) and equation (4) we derive an O$^{++}$/H$^+$ abundance 
of 7.62 dex and by adding the O$^{+}$ contribution a total O/H = 7.67 dex. 
The lower O/H value derived by Izotov et al. (2009) is not correct, their result probably has been affected by at least one of the following causes: a) that they did not consider the density dependence of the \ion{O}{2} lines, and b) that they did not use the proper branching ratio for the \ion{O}{2} lines.

\section{The Si/O ratio in \ion{H}{2} regions}\label{sec:Si/O} 

Si and O are produced by core collapse supernovae and this production is expected to be 
independent of the O/H ratio. By comparing the Si/O ratio in \ion{H}{2} regions with the 
Si/O ratio in B stars of the solar vicinity it is possible to estimate the fraction of Si 
in gas and dust phases in \ion{H}{2} regions.

\begin{deluxetable}{lr@{$\pm$}lr@{$\pm$}lr@{$\pm$}lr@{$\pm$}lr@{$\pm$}lr@{$\pm$}l}

\tablecaption{O/H visual and Si/O UV values
\label{tab:Si/O}}
\tablewidth{0pt}
\tabletypesize{\scriptsize}
\tablehead{
\colhead{Value}       &
\multicolumn{2}{c}{SMC N88A}    &
\multicolumn{2}{c}{NGC 2363}    &
\multicolumn{2}{c}{C1543+091}   &
\multicolumn{2}{c}{T$1214-277$}    &    
\multicolumn{2}{c}{SBS 0335$-$052 E} &
\multicolumn{2}{c}{I Zw 18}      
}
\startdata

O$_{gas}$/H; $t^2 = 0.00${\tablenotemark{a}}  
               & $ 8.06$&0.04 & $ 7.92$&0.04 & $ 7.76$&0.10 & $ 7.59$&0.04 & $ 7.31$&0.04 & $ 7.18$&0.04 \\
O$_{gas}$/H; $t^2 \neq 0.00${\tablenotemark{a}}  
               & $ 8.13$&0.10 & $ 8.00$&0.06 & $ 7.91$&0.10 & $ 7.74$&0.10 & $ 7.47$&0.04 & $ 7.29$&0.05 \\
Si$_{gas}$/O$_{gas}${\tablenotemark{b}} 
               & $-1.74$&0.12 & $-1.59$&0.14 & $-1.74$&0.14 & $-1.46$&0.28 & $-1.72$&0.20 & $-1.52$&0.22 \\
O$_{total}$/H; $t^2 \neq 0.00${\tablenotemark{c}}  
               & $ 8.23$&0.10 & $ 8.10$&0.07 & $ 8.01$&0.10 & $ 7.84$&0.10 & $ 7.51$&0.05 & $ 7.38$&0.05 \\
Si$_{gas}$/O$_{total}${\tablenotemark{c}} 
               & $-1.84$&0.12 & $-1.69$&0.14 & $-1.84$&0.14 & $-1.56$&0.28 & $-1.80$&0.20 & $-1.61$&0.22 \\
Si gas fraction{\tablenotemark{d}}   & 
\multicolumn{2}{c}{(23$\pm$6)\%} & \multicolumn{2}{c}{(32$\pm$10)\%} & \multicolumn{2}{c}{(23$\pm$7)\%} & \multicolumn{2}{c}{(44$+39\atop-20$)\%} & \multicolumn{2}{c}{(25$+15\atop-9$)\%} & \multicolumn{2}{c}{(39$+26\atop-16$)\%} \\

\enddata
\tablenotetext{a}{Garnett et al. (1995); Peimbert et al. (2000, 2007); this paper.}
\tablenotetext{b}{Garnett et al. (1995).}
\tablenotetext{c}{This paper, oxygen dust fraction from sections 4 and 8.}
\tablenotetext{d}{Under the assumption of a reference value of 
Si$_{total}$/O$_{total}$ = $-1.20$ dex.}

\end{deluxetable}

\begin{figure}
\plotone{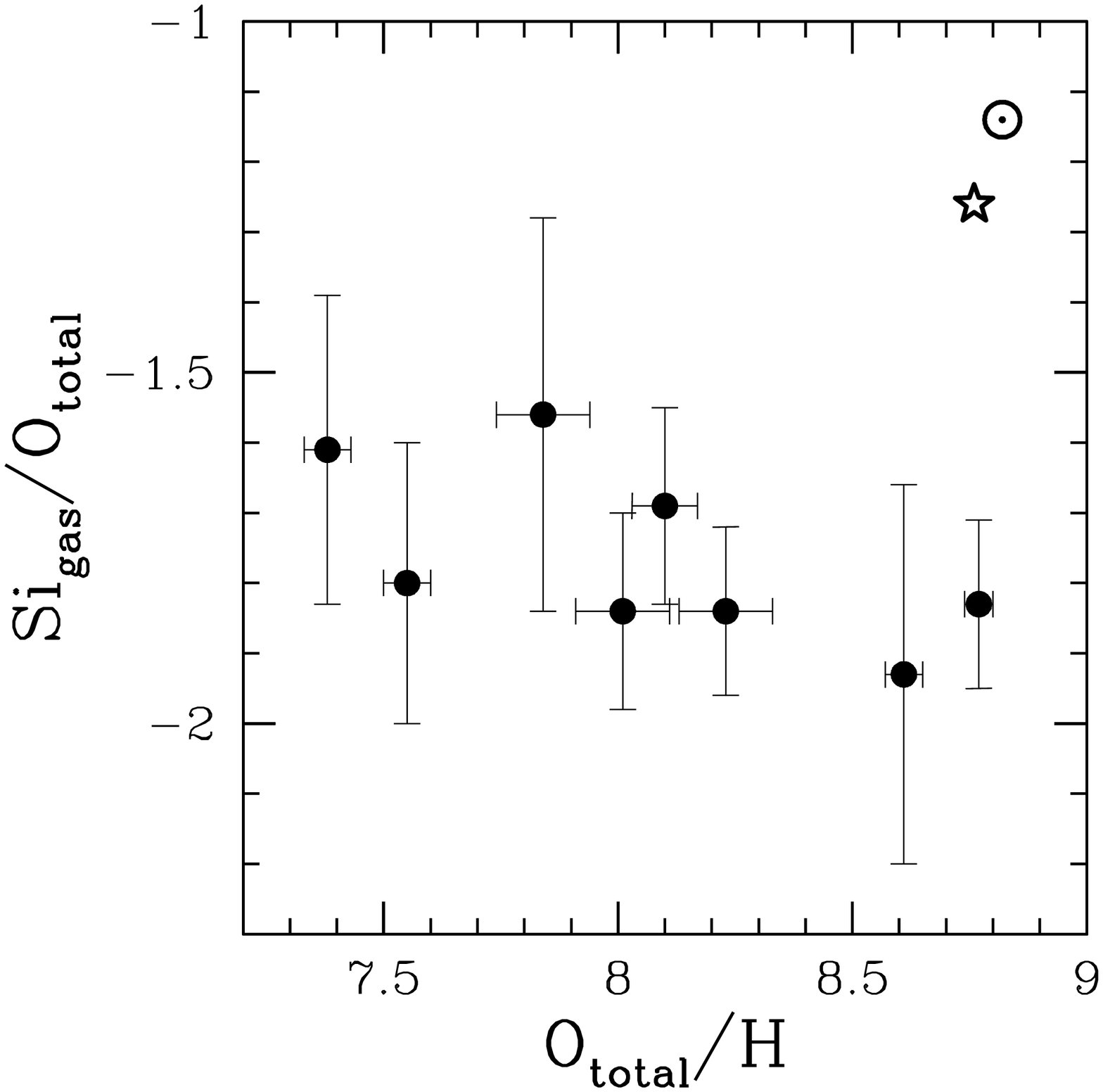}
\caption{Si$_{gas}$/O$_{total}$ versus O$_{total}$/H. The star represents the B stars of
the solar vicinity, the open circle with the central dot the protosolar + GCE value, and the filled 
circles the extragalactic \ion{H}{2}  regions ( see Tables 1 and 3).
\label{fig:Si/O}}
\end{figure}

Garnett et al. (1995) derived the Si/O ratio for eight Galactic and extragalactic 
\ion{H}{2} regions with gaseous O/H abundances in the 7.2 to 8.7 dex range and 
found that the Si/O ratio was almost constant for these objects. 
>From the Orion and 30 Doradus data of sections 2 and 3, and the data of the other 6 objects 
of Garnett et al. (see Table 3) we derived an average gaseous Si$_{gas}$/O$_{gas}$ = $-1.66\pm0.05$ dex. 
The Si$_{gas}$/O$_{total}$ versus O$_{total}$/H for the 8 \ion{H}{2} regions discussed in this paper 
are presented in Figure 5.

The reference value for Si/O amounts to $-1.20$ dex and is obtained 
from the protosolar + GCE Si/O value (Asplund et al. 2009; Chiappini et al. 
2003; Carigi \& Peimbert 2008, 2010) and the Si/O B stars values (Przybilla et al. 2008) 
that amount to $-1.14$ dex and $-1.26$ dex respectively.

Since the Orion nebula, the metal richest \ion{H}{2} region of the sample, 
has an oxygen depletion of 0.12 dex, and SBS~0335$-$052~E, one of the 2 metal poorest \ion{H}{2} 
regions of the sample, has an oxygen depletion of 0.08 dex, we can assume 
that most of the \ion{H}{2} regions have an intermediate depletion value. As a first 
approximation we will adopt 0.10 dex for the fraction of O trapped in dust grains. 
In section 7 this point will be discussed further and we will 
see that 0.10 dex is a good representative value for most \ion{H}{2} regions. 

>From the previous considerations we estimate  that the average ratio of Si$_{gas}$/O$_{total}$
amounts to $-1.76$ dex, from this estimate and the reference value of Si$_{total}$/O$_{total}$
of $-1.20$ dex we obtain that in \ion{H}{2} regions, about 28\% of the 
Si is in the gas phase and 72\% in the dust phase.

Note that the specific correction for oxygen in dust grains for the objects in Table 3 
(except SBS~0335$-$052~E) follow the recipe derived in section 8.

\section{The Fe/O ratio in \ion{H}{2} regions}\label{sec:Fe/O} 

\begin{figure}
\plotone{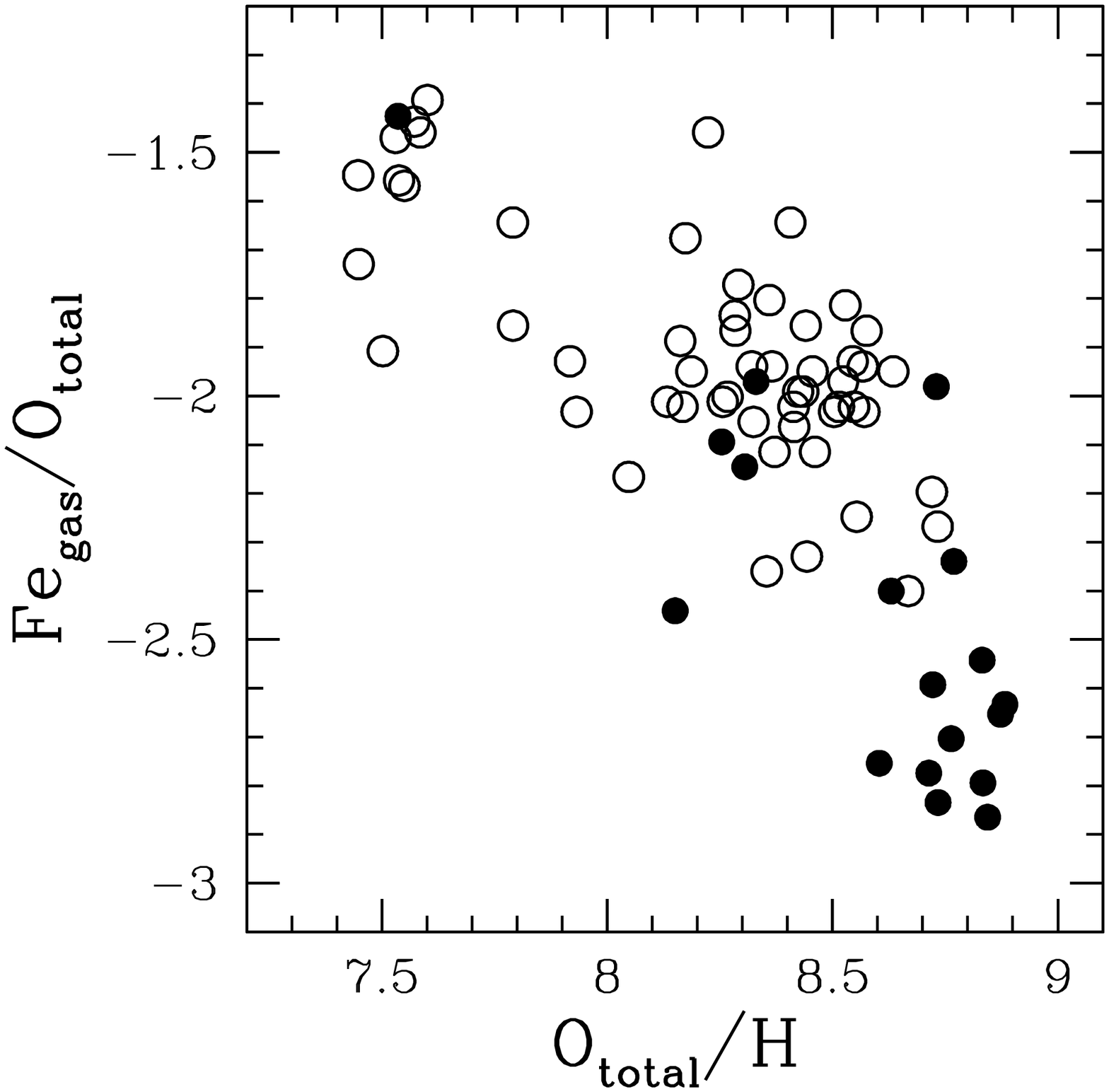}
\caption{Fe$_{gas}$/O$_{total}$  versus O$_{total}$/H, where to derive the O$_{total}$ from 
equation 2 we have assumed that the fraction of
Si in gas phase amounts to 28\%, that the fraction of Mg in gas phase is intermediate between
those of Si and Fe, and that the reference value for Fe$_{total}$/O$_{total}$ is $-1.20$ dex. The open
circles come from measurements by Izotov et al. (1997, 1998, 1999, 2004), where the O/H
values have been corrected 
for the presence of temperature variations, see text. The filled circles are from measurements by
Esteban et al. (2002, 2004, 2009), Garc\'{\i}a-Rojas et al. (2004, 2005, 2006, 2007), Peimbert 
et al. (2000), and this paper, where the O/H values are come from recombination lines.
\label{fig:Fe/O}}
\end{figure}

In Figure 6 we present the Fe/O versus O/H ratio compiled from many sources in the 
literature. The data comes from two different sets. In the first set the O/H determination 
comes from recombination line ratios which are independent of the temperature structure. 
In the second one the O/H ratios are derived in the traditional way, this 
depends on the ratio of collisionally exited to recombination lines
which is strongly dependent on the temperature structure of the \ion{H}{2} region 
(Peimbert, 1967; Peimbert \& Costero, 1969; Peimbert, M. \& Peimbert, A. 2010b).
 
To correct for the presence of temperature variations we have added 0.24 dex to 
the O/H ratio, this correction comes from the average value of 13 well observed 
extragalactic \ion{H}{2} regions (Esteban et al. 2002, 2009; L\'opez-S\'anchez 
et al. 2007). Note that the Fe/O ratio is almost independent of the temperature 
structure since it is determined from a ratio of collisionally excited lines that 
show a similar temperature dependence.

Part of the scatter in Figure 6 could be due to errors in the determinations of the 
gas-phase Fe/O ratios and part
to the different star formation histories of the different galaxies. The closer in time 
to a recent burst of star formation the lower the total Fe/O ratio in their ISM. In the 
solar vicinity at present all the O abundance is due to core collapse supernovae, while 
about 40\% of the Fe is due to core collapse supernovae and the other 60\% to Type Ia 
supernovae ( e. g. Pagel \& Tautvaisiene 1995, Pagel 2009). There is a time delay in 
the Fe formation relative to the O formation, and consequently the Fe/O ratio depends 
on the star formation rate, the initial stellar mass function, and the gas flows from 
and into the intergalactic medium. There are two well established Fe/O ratios from 
observations: the one when the Sun was formed, called the protosolar ratio that amounts 
to $-1.19$ dex (Asplund et al. 2009), and the present value in the solar vicinity based 
on observations of B stars that amounts to $-1.32$ dex (Przybilla et al. 2008); from the 
protosolar value plus the modification due to Galactic chemical evolution we find a value of
Fe/O = $-1.08$ dex for the present ISM (see Table 1). From the average of the previous 2 
values we obtain an Fe/O = $-1.20$ dex.

>From Figure 6 and under the assumption that the total Fe/O ratio amounts to $-1.20$ dex,
we obtain that for the Galactic and extragalactic \ion{H}{2} regions with abundances in 
the  8.3 dex $<$ O/H $<$ 8.8 dex range the average fraction of Fe in the gaseous-phase 
is about 10\%, with some of them reaching 3\%. For 
the extragalactic \ion{H}{2} regions with abundances in the  
7.8 dex $<$ O/H $<$ 8.3 dex range the fraction of Fe in the gaseous-phase is about 
25\%. For the extragalactic \ion{H}{2} regions in the 7.3 dex$<$ O/H $<$ 7.8 dex 
range the fraction of Fe in the gaseous phase is about 50\%.

\section{Discussion}\label{sec:discussion}

The Fe/O variation as a function of O/H implies that in the high metallicity high density
Galactic \ion{H}{2} regions about 97\% of the Fe atoms are embedded in dust grains. This
fraction diminishes with diminishing O/H and in the metal poorest extragalactic \ion{H}{2} regions,
characterized by high electron temperature and low electron density, about 40\% of the Fe
atoms are embedded in dust grains. The Si/O ratio is approximately constant from the O
richest to the O poorest \ion{H}{2} regions, implying that for the whole O/H observed range
the fraction of Si embedded in dust grains amounts to about $72 \pm 10\%$. A possible explanation for these two results is that dust grains have a solid core with about 72 \% of the Si 
and about 40\% of the Fe atoms present in the ISM, and that in addition they have a softer 
mantle with about 57\% of the Fe present in the ISM. The softer mantle is destroyed at higher 
temperatures and lower densities, while the solid core remains intact in all the \ion{H}{2} 
regions of the sample.

We have only the estimate of Mg$_{gas}$/H for the Orion nebula and for 30 Doradus. The 
30 Doradus results imply that Mg$_{gas}$/H  is similar to Si$_{gas}$/H and 
a lot larger than Fe$_{gas}$/H. For the Orion nebula the Mg$_{gas}$/H is intermediate 
between Si$_{gas}$/H and Fe$_{gas}$/H. Jenkins (2009) finds that in lines of 
sight dominated by neutral material the Mg and Si depletions are very similar 
and considerably smaller than those of Fe. From these results we will explore two 
possibilities for the \ion{H}{2} region set: a) that  Mg$_{gas}$/H is equal to Si$_{gas}$/H, 
and b) that Mg$_{gas}$/H is intermediate between Si$_{gas}$/H and Fe$_{gas}$/H.

\begin{deluxetable}{lccc}
\tablecaption{Depletion of O in dust grains (dex)
\label{tab:Depletions}}
\tablewidth{0pt}
\tablehead{
\colhead{Mg$_{gas}$/H  Assumption} &
\colhead{Low O/H\tablenotemark{a}} &
\colhead{Intermediate O/H\tablenotemark{b}} &
\colhead{High O/H\tablenotemark{c}} 
}
\startdata

Si$_{gas}$/H behavior\tablenotemark{d}                     & $0.087 \pm 0.004$ & $0.100 \pm 0.005$ & $0.105 \pm 0.006$ \\
Intermediate behavior\tablenotemark{de}  & $0.084 \pm 0.005$ & $0.101 \pm 0.006$ & $0.108 \pm 0.007$ \\
Adopted\tablenotemark{f}                                   &  $0.09 \pm 0.01$  &  $0.10 \pm 0.01 $ &  $0.11 \pm 0.01$  \\
\enddata
\tablenotetext{a}{$ 7.3 < O/H < 7.8$.}
\tablenotetext{b}{$ 7.8 < O/H < 8.3$.}
\tablenotetext{c}{$ 8.3 < O/H < 8.8$.}
\tablenotetext{d}{Standard reference values (Mg/O = $-1.15$ dex, Si/O = $-1.20$ dex, Fe/O = $-1.20$ dex).}
\tablenotetext{e}{Mg$_{gas}$/H = (Si$_{gas}$/H+Fe$_{gas}$/H) / 2.}
\tablenotetext{f}{Note that there is an additional 0.03 uncertainty in the absolute calibration, see text.}
\end{deluxetable}

We found that in the range of metallicities studied here there is a small but measurable trend
in the sense that the depletion of O in dust grains increases with metallicity. The limiting
values for the O depletion are: 0.08 dex for the O poorest objects, based 
on the observations of SBS~0335$-$052~E, one of the O poorest objects known; and 
0.12 dex for the Orion nebula one of the richest objects with a good determination. Note that the upper limit 
given by equation (1) is 0.14 dex if no Mg, Si, or Fe atoms remain the gas phase.

We decided to divide the sample in three O/H groups: a) 7.3 $<$ O/H $<$ 7.8 dex, 
b) 7.8 $<$ O/H $<$ 8.3 dex, and c) 8.3 dex $<$ O/H $<$ 8.8 dex. We took the 
Fe$_{gas}$/O$_{total}$ average of all the \ion{H}{2} regions in each interval
as representative of each group, and from equation (1) we computed 
Table 4. In the first row of Table 4 we present the O depletions, under the assumption 
that the ISM reference values 
are given by Mg/O = $-1.15$ dex, Si/O = $-1.20$ dex, and Fe/O = $-1.20$ dex, 
and that Mg$_{gas}$/H is equal to Si$_{gas}$/H; in the second row of Table 4 
we have assumed that Mg$_{gas}$/H is intermediate between to the Si$_{gas}$/H and Fe$_{gas}$/H.
Finally in the last row of Table 4 we present our adopted values.

So far we have not included the systematic error present in equation 1 due the depletion determinations
of the Orion nebula (our absolute calibrator). This systematic error is comparable to the trends 
that we have found, and makes any given absolute determination uncertain, but the differences 
derived between two objects are not affected by this error. The systematic error is about one fourth 
of the estimated O depletion and, including the possible errors in the reference values of Mg, Si, and Fe, translates to a final error of 0.03 dex for each of the 3 intervals discussed in this paper.

\section{Conclusions}\label{sec:conclusions}

{From} the \ion{Mg}{2} 3d-4f $\lambda$4481 recombination line intensities we have derived the Mg/H 
abundance ratio for the Orion nebula and for 30 Doradus. From this abundance ratio we 
have estimated  that in the Orion nebula and in 30 Doradus about 9\% and 27\% of the Mg atoms
are in the gas phase respectively, and that the rest of the Mg atoms are embedded in dust grains.

We have reduced the FORS raw data for SBS~0335$-$052~E and have measured four of the eight 
\ion{O}{2} lines of multiplet 1 and from the 
$I$(4639 + 4641 + 4649 + 51)/$I$(H$\beta)_{rec} = 0.0231 \pm 0.0111$ value we obtain that 
12 + log (O$^{++}$/H$^+$)$=  7.36{+0.17 \atop -0.28}$ and that the total O/H  
value amounts to $7.41{+0.17 \atop -0.28}$ dex.

We find that the gaseous Fe/H ratio has a typical value of about 6.25 dex with a dispersion of about 0.30 dex for  \ion{H}{2} regions with 
O/H in the 7.3 to 8.8 dex range. The almost constancy of the gas-phase Fe/H ratio 
reflects the efficiency of the processes of dust formation and dust destruction. It 
probably implies that there is a minimum threshold for dust formation given by a 
gas-phase Fe/H ratio of about 5.8 dex.

Based on the Mg/O, Si/O, and Fe/O abundances of Galactic and extragalactic \ion{H}{2}
regions we estimate that: for the 8.3 dex $<$ O/H $<$ 8.8 dex range the fraction of 
O atoms trapped by dust grains amounts to $0.11 \pm 0.03$ dex, for the 
7.8 $<$ O/H $<$ 8.3 dex range amounts to $0.10 \pm 0.03$ dex, and for the 
7.3 $<$ O/H $<$ 7.8 dex range amounts to $0.09 \pm 0.03$ dex. Note that if one of the adopted 
reference values for Mg, S, and Fe  were increased or decreased by 0.1 dex 
the O depletions mentioned above change by about 0.002 dex.

We also consider that the \ion{H}{2} region abundances derived from the $T$(4363/5007) method 
underestimate the O/H ratio by about 0.15 to 0.35 dex (e. g. Peimbert, M. \& Peimbert, 
A. 2010b, and references therein). This result together with the fraction of O atoms embedded 
in dust grains imply that to obtain the gas-phase plus the dust-phase O/H value it is necessary 
to increase the O/H gas values derived from the $T$(4363/5007) method by about 0.25 to 0.45 dex.

\acknowledgments

We would like to acknowledge the referee for a careful review of the manuscript and for many
excellent suggestions. This work was partly supported by the grants PAPIIT IN123309 from DGAPA 
(UNAM, Mexico) and CONACyT 46904 (Mexico).


\begin{thebibliography}

\bibitem[Asplund et al. (2009)]{asp09}
Asplund, M., Grevesse, N., Sauval, A. J., \&  Scott, P. 2009,  
Ann. Rev. A. \& Ap., 47, 481

\bibitem[Baldwin et al. (1991)]{bal91}
Baldwin, J. A., Ferland, G. J., Martin, P. G., Corbin, M. R., Cota, S. A., Peterson, B. M., 
\& Slettebak, A. 1991,
\apj, 374, 580

\bibitem[Bastin \& Storey (2006)]{bas06} 
Bastin, R. J. \& Storey, P. J. 2006,
in Planetary Nebulae in our Galaxy and Beyond, 
IAU Symposium 234, M. J. Barlow \& R. H. M\'endez (eds.), (Cambridge:
Cambridge Univ. Press), p. 369

\bibitem[Cannon et al. (2002)]{can02}
Cannon, J. M., Skillman, E. D., Garnett, D. R., \& Dufour, R. J. 2002, 
\apj, 565, 931

\bibitem[Carigi (1996)]{car96}
Carigi, L. 1996, 
RevMexAA, 32, 179

\bibitem[Carigi \& Peimbert (2008)]{car08}
Carigi, L. \& Peimbert, M. 2008, 
RevMexAA, 44, 341

\bibitem[Carigi \& Peimbert (2010)]{car10}
Carigi, L. \&  Peimbert, M. 2010,
\apj, submitted, arXiv1004.0756

\bibitem[Chiappini et al. (2003)]{chi03}
Chiappini, C., Romano, D., \& Matteucci, F. 2003,
\mnras, 339, 63

\bibitem[Deharveng et al. (2000)]{deh00} 
Deharveng, L., Pe\~na, M., Caplan, J., \& Costero, R. 2000, 
\mnras, 311, 329

\bibitem[Esteban et al. (2009)]{est09}
Esteban, C., Bresolin, F., Peimbert, M., Garc\'{\i}a-Rojas, J., Peimbert, A., \& Mesa-Delgado, A. 2009, 
\apj, 700, 654

\bibitem[Esteban et al. (2005)]{est05}
Esteban, C., Garc\'{\i}a-Rojas, J., Peimbert, M., Peimbert, A., Ruiz, M. T., 
Rodr\'{\i}guez, M., \& Carigi, L. 2005, 
\apj, 618, L95

\bibitem[Esteban et al. (2004)]{est04} 
Esteban, C., Peimbert, M., Garc\'{\i}a-Rojas, J., Ruiz, M. T., Peimbert, A., \& Rodr\'{\i}guez, M. 2004,
\mnras, 355, 229

\bibitem[Esteban et al.(1998)]{est98} 
Esteban, C., Peimbert, M., Torres-Peimbert, S., \& Escalante, V. 1998,
\mnras, 295, 401

\bibitem[Esteban et al. (2002)]{est02}
Esteban, C., Peimbert, M., Torres-Peimbert, S., \& Rodr\'{\i}guez, M. 2002,
\apj, 581, 241

\bibitem[Fadeyev(1988)]{fad88}  
Fadeyev, Y. 1988,
in Atmospheric Diagnostics of Stellar Evolution, ed. K. Nomoto (Berlin: Springer-Verlag), 533 

 \bibitem[Garc\'{\i}a-Rojas et al. (2005)]{gar05}
Garc\'{\i}a-Rojas, J., Esteban, C., Peimbert, A., Peimbert, M., Rodr\'{\i}guez, M., \& Ruiz, M. T. 2005,
\mnras, 362, 301

\bibitem[Garc\'{\i}a-Rojas et al. (2007)]{gar07}
Garc\'{\i}a-Rojas, J., Esteban, C., Peimbert, A., Rodr\'{\i}guez, M., 
Peimbert, M., \& Ruiz, M. T. 2007,
RevMexAA, 43, 3

\bibitem[Garc\'{\i}a-Rojas et al. (2006)]{gar06}
Garc\'{\i}a-Rojas, J., Esteban, C., Peimbert, M., Costado, M. T., Rodr\'{\i}guez, M., Peimbert, A., \& Ruiz, M. T. 2006,
\mnras, 368, 253

\bibitem[Garc\'{\i}a-Rojas et al. (2004)]{gar04} 
Garc\'{\i}a-Rojas, J., Esteban, C., Peimbert, M., Rodr\'{\i}guez, M., Ruiz, M. T., \& Peimbert, A. 2004,
\apjs, 153, 501

\bibitem[Garnett et al. (1995)]{gar95}
Garnett, D. R., Dufour, R. J., Peimbert, M., Torres-Peimbert, S., Shields, G. A., 
Skillman, E. D., Terlevich, E., \& Terlevich, R. J. 1995,
\apj, 449, L77
	
\bibitem[Izotov et al. (1999)]{izo99}
Izotov, Y. I., Chaffee, F. H., Foltz, C. B., Green, R. F., Guseva, N. G., \& Thuan, T. X. 1999,
\apj, 527, 757

\bibitem[Izotov et al. (2009)]{izo09}
Izotov, Y. I., Guseva, N. G., Fricke, K. J., \& Papaderos, P. 2009,
\aap, 503, 61	
	
\bibitem[Izotov et al. (2006)]{izo06}
Izotov, Y. I., Stasi\'nska, G., Meynet, G., Guseva, N. G., \& Thuan, T. X. 2006,
\aap, 448, 955

\bibitem[Izotov \& Thuan (1998)]{izo98}
Izotov, Y. I. \& Thuan, T. X. 1998,
\apj, 500, 188

\bibitem[Izotov \& Thuan (2004)]{izo04}	
Izotov, Y. I. \&  Thuan, T. X. 2004,
\apj, 602, 200

\bibitem[Izotov et al. (1997)]{izo97}
Izotov, Y. I., Thuan, T. X., \& Lipovetsky, V. A. 1997,
\apjs, 108, 1

\bibitem[Jenkins (2009)]{jen09}	
Jenkins, E. B. 2009,
\apj, 700, 1299

\bibitem[Keenan et al. (2001)]{kee01}
Keenan, F. P., Aller, L. H., Ryans, R. S. I., \& Hyung, S. 2001,
P. Natl. Acad. Sci. USA, 98, 9476

\bibitem[Liu et al. (2004)]{liu04}
Liu, Y., Liu, X.-W., Barlow, M. J., \& Luo, S.-G 2004,
\mnras, 353, 1251

\bibitem[L\'opez-S\'anchez et al. (2007)]{lop07}
L\'opez-S\'anchez, A. R., Esteban, C., Garc\'{\i}a-Rojas, J., Peimbert, M.,
\& Rodr\'{\i}guez, M. 2007,
\apj, 656, 168

\bibitem[Luridiana et al. (2003)]{lur03}
Luridiana, V., Peimbert, A., Peimbert, M. \&  Cervi\~no, M. 2003,
\apj, 592, 846

\bibitem[Mesa-Delgado et al. (2009)]{mes09}
Mesa-Delgado, A., Esteban, C., Garc\'{\i}a-Rojas, J., Luridiana, V., Bautista, M., 
Rodr\'{\i}guez, M., L\'opez-Mart\'{\i}n, L., \& Peimbert, M. 2009, 
\mnras, 395, 855

\bibitem[Nuth \& Hecht (1990)]{nut90}	
Nuth, J. A. \& Hecht, J. H. 1990,
Ap\&SS, 163, 79

\bibitem[Ossenkopf et al. (1992)]{oss92}
Ossenkopf, V., Henning, Th., \& Mathis, J. S. 1992,
\aap, 261, 567

\bibitem[Pagel (2009)]{pag09}
Pagel, B. E. J. 2009, in Nucleosynthesis and Chemical Evolution of Galaxies,
 Second Edition, (Cambridge: Cambridge Univ. Press)

\bibitem[Pagel \& Tautvaisiene (1995)]{pag95}
Pagel, B. E. J. \& Tautvaisiene, G. 1995,
\mnras, 276, 505

\bibitem[Peimbert (2003)]{pei03} 
Peimbert, A. 2003,
\apj, 584, 735

\bibitem[Peimbert \& Peimbert (2005a)]{pei05a}
Peimbert, A. \& Peimbert, M. 2005a, 
RevMexAA, SC, 23, 9

\bibitem[Peimbert \& Peimbert (2010a)]{pei10a}
Peimbert, A. \& Peimbert, M. 2010a,
in Light Elements in the Universe, IAU Symposium 268, C. Charbonnel, M. Tosi, 
F. Primas, \& C. Chiappini, (eds.), (Cambridge: Cambridge Univ. Press), 185

\bibitem[Peimbert et al. (2005b)]{pei05b}
Peimbert, A., Peimbert, M., \&  Ruiz, M. T, 2005b,
\apj, 634, 1056

\bibitem[Peimbert (1967)]{pei67} 
Peimbert, M. 1967, 
\apj, 150, 825

\bibitem[Peimbert \& Costero (1969)]{pei69} 
Peimbert, M. \& Costero, R. 1969,
Bol. Obs. Tonantzintla y Tacubaya, 5, 3

\bibitem[Peimbert et al. (2007)]{pei07}
Peimbert, M., Luridiana, V., \& Peimbert, A. 2007,
\apj, 666, 636

\bibitem[Peimbert \& Peimbert (2010b)]{pei10b}
Peimbert, M. \& Peimbert, A. 2010b,
RevMexAA, SC, arXiv: 0912.3781, in press

\bibitem[Peimbert et al. (2000)]{pei00}
Peimbert, M., Peimbert, A., \& Ruiz M. T. 2000,
\apj, 541, 688 

\bibitem[Perinotto \& Patriarchi (1980)]{per80}
Perinotto, M. \& Patriarchi, P. 1980,
\apj, 235, L13

\bibitem[Pilyugin et al. (2003)]{pil03} 
Pilyugin, L. S., Ferrini, F., \& Shkvarun, R. V. 2003, 
\aap, 401, 557

\bibitem[Przybilla, et al. (2008)]{prz08}
Przybilla, N., Nieva, M. F., \& Butler, K. 2008, 
\apj, 688, L103

\bibitem[Rodr\'{\i}guez \& Rubin (2005)]{rod05}
Rodr\'{\i}guez, M. \& Rubin, R. H. 2005,
\apj, 626, 900

\bibitem[Ruiz et al. (2003)]{rui03}
Ruiz, M. T., Peimbert, A., Peimbert, M., \& Esteban, C.  2003,
\apj, 595, 247

\bibitem[Simón-D\'{\i}az(2010)]{sim10}
Sim\'on-D\'{\i}az, S. 2010,
\aap, 510, 22

\bibitem[Storey(1994)]{sto94}
Storey, P. J., 1994,
\aap, 282, 999

\bibitem[Storey \& Hummer(1995)]{sto95}
Storey, P. J. \& Hummer, D. G. 1995,
\mnras, 272, 41

\bibitem[Thuan et al. (1997)]{thu97}
Thuan, T. X., Izotov, Y. I., \& Lipovetsky, V. A. 1997, 
\apj, 477, 661

\bibitem[Tsamis \& P\'equignot (2005)]{tsa05}
Tsamis, Y. G. \& P\'equignot, D. 2005,
\mnras, 364, 687

\end{thebibliography}
\end{document}